\documentclass[aip,amsmath,amssymb,reprint]{revtex4-2}
\usepackage{graphicx}
\usepackage{dcolumn}
\usepackage{bm}
\usepackage[utf8]{inputenc}
\usepackage[T1]{fontenc}
\usepackage{mathptmx}
\usepackage{etoolbox}
\usepackage[separate-uncertainty]{siunitx}
\usepackage{graphicx}

\makeatletter
\def\@email#1#2{%
 \endgroup
 \patchcmd{\titleblock@produce}
  {\frontmatter@RRAPformat}
  {\frontmatter@RRAPformat{\produce@RRAP{*#1\href{mailto:#2}{#2}}}\frontmatter@RRAPformat}
  {}{}
}%
\makeatother

\renewcommand{\vec}[1]{{\ensuremath{\mathbf{\boldsymbol#1}}}}

\newcommand{\DOT}[2]{\ensuremath{\vec{ #1 }\cdot\vec{ #2 }}}
\newcommand{\CROSS}[2]{\ensuremath{\vec{ #1 }\times\vec{ #2 }}}
\newcommand{\DIV}[1]{\ensuremath{\DOT{\nabla}{ #1 }}}
\newcommand{\CURL}[1]{\ensuremath{\CROSS{\nabla}{ #1 }}}
\newcommand{\GRAD}[0]{\ensuremath{\vec{\nabla}}}
\newcommand{\AVE}[1]{\ensuremath{\left\langle #1 \right\rangle}}
\newcommand{\NORM}[1]{\ensuremath{\lVert #1 \rVert}}

\begin{document}

\title{Complex Dynamics of an Acoustically Levitated Fluid Droplet Captured by a Low-Order Immersed Boundary Method}

\author{Jacqueline B. Sustiel}
\author{David G. Grier}
\email{david.grier@nyu.edu}
\affiliation{Department of Physics and Center for Soft Matter Research, New York University, New York, NY 10003, USA}


\begin{abstract}
We present a novel immersed boundary method that implements acoustic perturbation theory to model an acoustically levitated droplet. Instead of resolving sound waves numerically, our hybrid method solves acoustic scattering semi-analytically and models the corresponding time-averaged acoustic forces on the droplet. This framework allows the droplet to be simulated on inertial timescales of interest, and thereby admit a much larger time resolution than traditional compressible flow solvers. To benchmark this technique and demonstrate its utility, we implement the hybrid IBM for a single droplet in a standing wave. Simulated droplet shape deformations and streaming profile agree with theoretical predictions. Our simulations also yield new insights on the streaming profiles for elliptical droplets, for which a comprehensive analytic solution does not exist.
\end{abstract}
\maketitle

\section{Introduction}

Acoustic manipulation uses the forces exerted
by sound waves to lift objects against gravity
and to move them along planned trajectories
in three dimensions \cite{santesson2004airborne,andrade2018review,mu2019mass,marzo2019holographic,abdelaziz2021ultrasonic}.
The theory of acoustic manipulation is well-developed for
solid objects in inviscid fluids \cite{silva2018acoustic}.
The behavior of deformable objects, however, poses more of a
challenge.
Acoustic scattering and sound-mediated shape changes occur
on dramatically different time scales, making their interplay difficult to model self-consistently \cite{karlsen2015forces,wang2020immersed}.
The standard formulation of
sound-mediated forces and flows
is based on second-order perturbation theory.
The first-order pressure and velocity fields,
$p_1(\vec{r}, t)$ and $\vec{u}_1(\vec{r}, t)$,
respectively, describe incident
sound waves together with waves
scattered by particles.
These first-order fields oscillate harmonically
at the acoustic frequency, $\omega$, and vanish on average.
Their interference, however, gives rise to
steady acoustic radiation forces (ARF) \cite{bruus2012acoustofluidics7}
and streaming flows
\cite{nyborg1958acoustic,qi1993effect,lee1990outer,westervelt1953theory,baasch2020acoustic} at second order in the amplitude of
the sound wave.
This wave-matter interaction has been formulated
analytically for highly symmetric scatterers
\cite{marston1980shape,bruus2012acoustofluidics2,mitri2017axial,silva2018acoustic,baasch2020acoustic}
and semi-analytically for scatterers of fixed arbitrary shape \cite{waterman2009t}.
Even numerical solutions are challenging,
however, for
scatterers whose shape can change.

Here, we introduce an efficient computational framework
based on the Immersed Boundary Method (IBM)
\cite{peskin1972flow,peskin2002immersed}
that accurately predicts the dynamics and
shape evolution of fluid droplets in acoustic force landscapes,
as well as the streaming flows around them.
The IBM naturally accommodates the moving boundary of a deformable droplet and couples
it to flows in the fluid medium.
We then couple
this dynamical system
semi-analytically
to an acoustic pressure wave
by projecting the time-averaged acoustic force at the deformed droplet's surface onto the incompressible flow
around an equivalent sphere.
This framework models acoustic levitation
substantially more efficiently than existing
compressible-flow solvers
by inherently accommodating both fast and
slow processes.
We validate this hybrid numerical framework by studying
shape deformations and flow fields around fluid droplets
in an acoustic levitator.
This study identifies a
dynamical transition between dipolar and quadrupolar
streaming flows
that may explain anomalous collective
phenomena in experimental studies
of acoustically levitated droplets
\cite{abdelaziz2021ultrasonic}
and can be useful for designing
microfluidic devices \cite{maramizonouz2021numerical}.

Section~\ref{sec:ibm}
reviews the standard IBM
\cite{peskin2002immersed} along with
extensions that incorporate
the mass and surface tension of a
fluid droplet.
Acoustic forces and streaming flows
are derived in Sec.~\ref{sec:acoustic}
from acoustic perturbation theory.
Section~\ref{sec:implementation}
then
The outcome compares well with results of the best available multiscale computational
studies at a fraction of the computational cost.

\begin{figure}
\centering
\includegraphics[width=0.9\columnwidth]{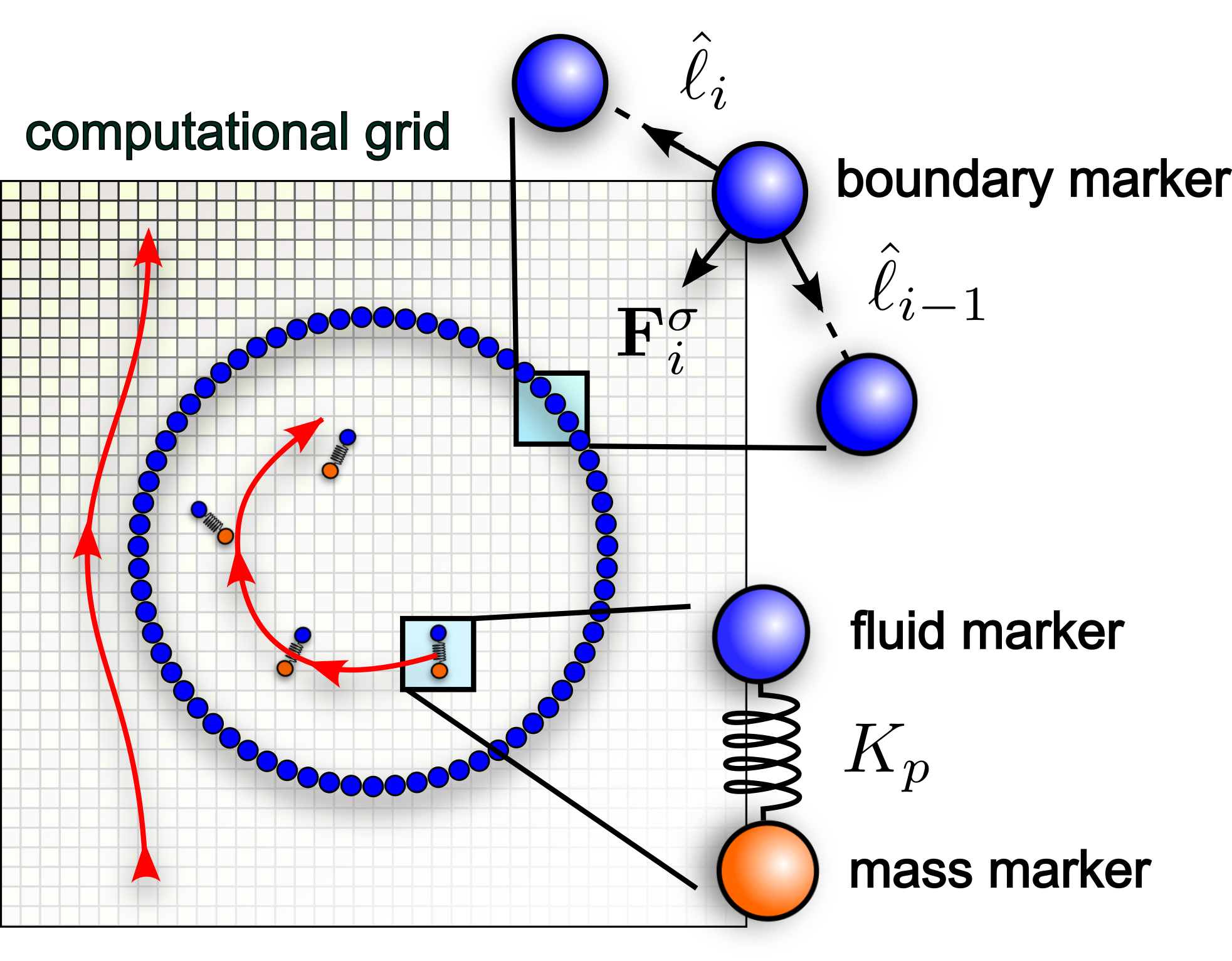}
\caption{Schematic overview of the immersed boundary method applied to a fluid droplet driven by an acoustic pressure field.}
\label{fig:schematic}
\end{figure}

\section{Immersed Boundary Method}
\label{sec:ibm}
The Immersed Boundary Method was first introduced in 1972 to study blood flow in the heart \cite{peskin1972flow}, and has since evolved into a general method for solving
fluid-structure interaction problems \cite{peskin2002immersed}.
Modeling fluid droplets requires a variant of this technique called
the penalty Immersed Boundary Method that incorporates the
droplets' mass and surface tension \cite{kim2007penalty, kim2016penalty}.
The present work further extends pentalty IBM
by further incorporating the forces exerted
by an acoustic wave propagating through the fluid medium.

\subsection{The standard IBM}
\label{sec:standardibm}

The standard formulation of the IBM
\cite{peskin2002immersed} applies to a droplet
embedded in a homogeneous incompressible fluid of
density $\rho$ and dynamic viscosity $\eta$.
The droplet's surface, $\vec{X}(\vec{q})$, is parameterized by curvilinear coordinates, $\vec{q}$,
and is discretized into
a set of markers,
$\{\vec{X}_i\}$, at fixed material points along the boundary,
as shown schematically in Fig.~\ref{fig:schematic}.

The fluid flow, $\vec{u}(\vec{x}_j)$,
and pressure, $p(\vec{x}_j)$,
are defined on a Cartesian grid, $\vec{x}_j$, with uniform spacing $h$,
and are governed by
the viscous, incompressible Navier-Stokes equations,
\begin{subequations}
\label{eq:incNS}
\begin{gather}
    \rho \, (\partial_t
    + \vec{u}\cdot\vec{\nabla}) \, \vec{u}
    = -\vec{\nabla}p
    + \eta\nabla^2\vec{u} + \vec{f} \\
    \vec{\nabla}\cdot\vec{u}
    = 0 .
\end{gather}
\end{subequations}
By Newton's third law, the force on the fluid element
at $\vec{x}_j$,
\begin{equation}
    \vec{f}(\vec{x}_j, t)
     =
    - \int
    \vec{F}(\vec{q}, t) \,
    \delta_h(\NORM{\vec{x}_j - \vec{X}(\vec{q}, t)}) \,
    d\vec{q} ,
    \label{eq:IB1}
\end{equation}
is equal and opposite to
the local force on the boundary, $\vec{F}(\vec{q}, t)$, due to
surface tension and
the influence of external fields.
The kernel, $\delta_h$, interpolates
over the markers, $\vec{X}_i$, in the neighborhood of a discrete gridpoint, $\vec{x}_j$.
Likewise, each massless boundary marker
is advected by the fluid at a velocity
\begin{equation}
    \vec{U}_i
     =
    \int
    \vec{u}(\vec{x}_j, t) \,
    \delta_h(\NORM{\vec{x}_j - \vec{X}_i})
    \, d\vec{x}_j
    \label{eq:IB2}
\end{equation}
that is interpolated from the computational grid.

\begin{figure}
    \centering
    \includegraphics[width=\columnwidth]{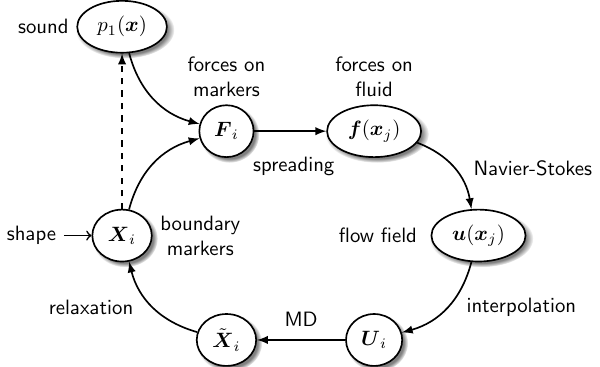}
    \caption{Immersed Boundary Method for acoustic manipulation of
    deformable objects.}
    \label{fig:ibm}
\end{figure}

The IBM defined by Eqs.~\eqref{eq:incNS}-\eqref{eq:IB2}
generally proceeds in a cycle, as depicted in Fig.~\ref{fig:ibm}.
Given an initial droplet shape, $\{\vec{X}_i\}$,
the forces, $\{\vec{F}_i\}$, on the
boundary markers are calculated from the
combined influences of surface tension and external forces.
The complementary force field acting on the fluid, $\vec{f}(\vec{x}_j, t)$,
is then computed with Eq.~\eqref{eq:IB1}.
The Navier-Stokes equations, Eq.~\eqref{eq:incNS}, then are
solved numerically \cite{peskin1972flow, peskin2002immersed} to obtain the updated fluid velocity,
$\vec{u}(\vec{x}_j, t)$.
Finally, Eq.~\eqref{eq:IB2} is used to determine the no-slip motion of the boundary, $\partial_t\vec{X}=\vec{U}$, and the procedure is repeated for the updated boundary configuration, $\{\vec{X}_j\}$.

In this work, we implement the same numerical scheme first introduced by Peskin \cite{peskin1972flow}, using the spacial and temporal discretizations and four-point discrete delta function described in Sections 4-7 of Ref.~\cite{peskin2002immersed}. To improve volume conservation, we calculate the fluid fields on a staggered, marker-and-cell grid \cite{fai2013varvisc}.

\subsection{Penalty IBM for Massive Droplets}
\label{sec:penaltyibm}

The standard IBM cannot account for multiphase flows where
properties of the droplet, such as density and viscoelasticity,
differ from those of the surrounding medium
\cite{fai2013varvisc}.
For simplicity, we focus on the droplet's
density contrast relative to the medium.
To incorporate the droplet's excess mass, we implement an extension of the IBM
based on the penalty Immersed Boundary Method
\cite{kim2007penalty,kim2008numerical}.

The penalty IBM models the density of the
droplet interior by introducing a a set of
massless tracer particles at locations
$\{\vec{X}_i^m\}$
whose trajectories trivially satisfy Eqs.~\eqref{eq:incNS}-\eqref{eq:IB2}.
As shown in Fig.~\ref{fig:schematic},
each of these fluid markers is coupled to a corresponding \emph{mass marker} at
position $\vec{Y}_i^m$ by a spring of stiffness $K_P$.
The $i$-th fluid marker experiences a force,
\begin{equation}
    \vec{F}_i^m = K_P\left(\vec{Y}_i^m-\vec{X}_i^m\right),
\end{equation}
that accelerates it according to Newton's second law,
\begin{equation}
    M \, \partial_t^2 \vec{Y}_m = -\vec{F}_i^m -Mg\hat{z},
\end{equation}
where $M$ is the marker's mass and $g$ is the acceleration due to gravity.

In theory, the fluid markers and mass markers ought to coincide in the limit of large $K_P$. In practice, however, numerical instability arises when $K_P$ is too large. Consequently, $K_P$ is chosen phenomenologically to maintain pair separations smaller than the
computational grid size, typically at $h/10$.

\subsection{Surface Tension}
\label{sec:surfacetension}

The internal boundary stress on a marker at $\vec{X}_i$ is often formulated as a
restoring force, $\vec{F}_i = - \delta_{\vec{X}_j} E$,
that minimizes the interfacial energy,
$E[\{\vec{X}_j\}]$.
When this approach is used to model surface tension,
the interfacial energy is proportional to the droplet's
total surface area.
In two dimensions, it
is proportional to the length of the interface,
\begin{equation}
    E[\{\vec{X}_j\}]
    =
    \sigma A
    =
    \sigma\sum_{j=1}^{N_{ib}-1}
    \NORM{\vec{\ell}_j},
\end{equation}
where
$\vec{\ell}_j \equiv \vec{X}_{j+1}-\vec{X}_j$, and $\sigma$ is the interfacial tension.
The force of surface tension
acting on the marker is then
\begin{equation}
    \label{eq:surfacetension}
    \vec{F}^{\sigma}_i
    \Big(\sigma \sum_{j=1}^{N_{ib}-1}||\vec{\ell}_j||\Big)
    = - \sigma(\hat{\ell}_i - \hat{\ell}_{i-1}).
\end{equation}
Equation~\eqref{eq:surfacetension} correctly
models surface tension if the boundary
markers are uniformly distributed around the droplet's boundary.
To this end, the markers are redistributed tangentially along the boundary at each time step.

Surface tension exerts a normal on the droplet surface to minimize local curvature.
Consequently, Eq.~\eqref{eq:surfacetension} also defines the surface normal, $\hat{\mathbf{n}}_i = \vec{F}_i^\sigma/\NORM{\vec{F}_i^\sigma}$.

\subsection{Benchmark: Sedimenting Droplet}
\label{sec:sedimentingdroplet}

\begin{figure}
    \centering
    \includegraphics[height=3.2in]{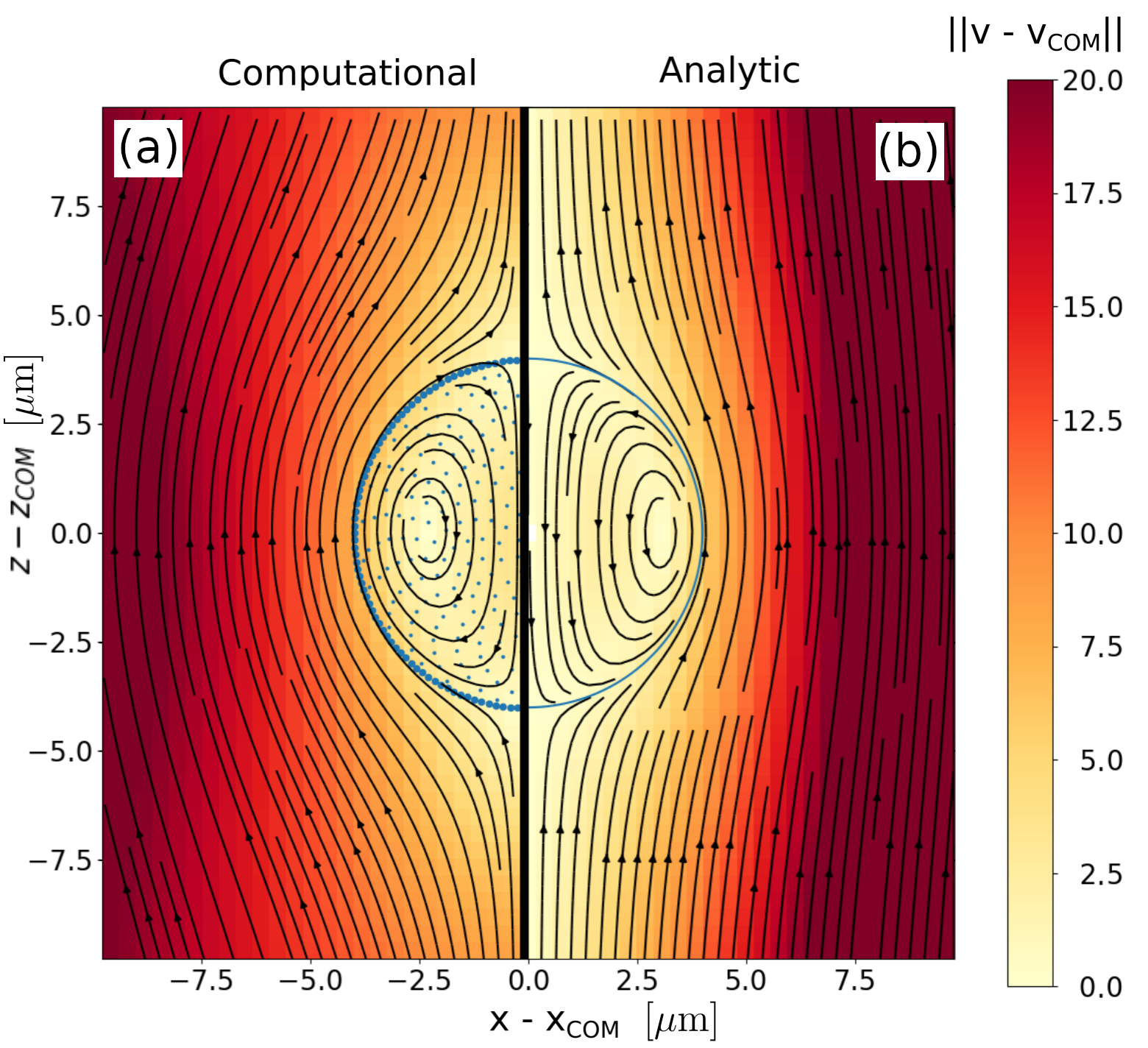}
    \caption{Streamlines within and around a \SI{4}{\um} droplet falling through an unbounded fluid at its terminal velocity, $\approx\SI{20}{\um\per\s}$. We compare the analytic stream function from Happel and Brenner \cite{happel1983low} (left) to simulations using a penalty immersed boundary method for the
    droplet's excess mass (right).
    }
    \label{fig:streamfall}
\end{figure}

To demonstrate and benchmark our implementation of the IBM, we model the sedimentation of a droplet in an unbounded fluid \cite{kim2016penalty}.
Figure~\ref{fig:streamfall} presents a cross-section
of the flow around and within a droplet of
radius $R = \SI{4}{\um}$ and
buoyant density
$\rho_p - \rho_0 = \SI{0.18}{\gram\per\cubic\cm}$,
which corresponds to silicone oil
sedimenting in water \cite{abdelaziz2021ultrasonic}.
For simplicity, we assume the fluid droplet
has the same viscosity as the surrounding
medium.
The plot shows streamlines of the flow in the
co-moving frame and is colored by the local
flow speed.
The data in Fig.~\ref{fig:streamfall}(a)
are computed with the IBM in a
\qtyproduct{20x20}{\um} periodic
domain.
They are compared side-by-side
with the analytic result for a spherical droplet \cite{happel1983low}
in Fig.~\ref{fig:streamfall}(b).
This solution for the two-phase flow is
reproduced in Appendix~\ref{sec:sedimenting}.

Slight differences between the two flow fields
can be attributed to the influence of
periodic boundary conditions on the simulations
and to the implicit assumption of infinite surface
tension in the analytic model.

\section{Acoustics}
\label{sec:acoustic}

The standard IBM relies on
the incompressible Navier-Stokes equations, Eq.~\eqref{eq:incNS},
and therefore does not accommodate sound waves.
Describing the fields within an acoustic levitator
requires the compressible Navier-Stokes equations,
\begin{subequations}
    \label{eq:compressiblenavierstokes}
\begin{align}
    \partial_t\rho
    &=
    -\vec{\nabla} \cdot (\rho\vec{u})\\
    \rho \, (\partial_t + \vec{u}\cdot\vec{\nabla}) \, \vec{u}
    &=
    -\vec{\nabla}p + \eta \, \nabla^2\vec{u} + \beta\eta \, \vec{\nabla}(\vec{\nabla}\cdot\vec{u}) ,
\end{align}
\end{subequations}
where $\beta = \xi/\eta + 1/3$
incorporates the fluid's volume viscosity,
$\xi$
\cite{baasch2020acoustic, settnes2012forces}.

In the absence of sound waves, the fluid medium is
quiescent, $\vec{u} = 0$, and has uniform pressure $p_0$, and uniform density, $\rho_0$.
An acoustic levitator projects a time-harmonic pressure wave,
$p_\text{inc}(\vec{x}, t) = p_\text{inc}(\vec{x})\exp(-i\omega t)$,
into the fluid.
The droplet scatters a portion of $p_\text{inc}$ to
create a scattered wave,
$p_\text{scat}(\vec{x}, t)
=
p_\text{scat}(\vec{x}, \{\vec{X}_j\}) \, \exp(-i\omega t)$,
that depends on the position and shape of the droplet.
The incident and scattered waves together
cause small pressure fluctuations,
$p_1(\vec{x}) = p_\text{inc}(\vec{x}) + p_\text{scat}(\vec{x})$, about $p_0$.
This first-order pressure wave is associated with
a first-order density wave, $\rho_1(\vec{x})$, and a
first-order velocity field, $\vec{u}_1(\vec{x})$,
that are obtained by expanding Eq.~\eqref{eq:compressiblenavierstokes}
to first order in the fields.

Acoustic levitators typically operate at
such high frequencies that the period of
$p_1(\vec{x}, t)$ is much shorter than
the viscous and inertial time scales that
govern droplet dynamics and streaming flows.
Those comparatively slow processes emerge
as averages over multiple acoustic cycles.
Because the first-order fields are harmonic and therefore vanish on average, standard acoustic perturbation theory \cite{bruus2012acoustofluidics7}
expands the density,
pressure and velocity fields to second order:
\begin{subequations}
\label{eq:pert}
\begin{align}
    \rho
    & =
    \rho_0 + \rho_1 + \rho_2 \\
    p
    & =
    p_0 + p_1 + p_2 \\
    \vec{u}
    & =
    \vec{u}_1 + \vec{u}_2 .
\end{align}
\end{subequations}
The second-order fields, $\rho_2$, $p_2$ and $\vec{u}_2$, do not vanish on average, and therefore describe steady dynamics that persist on hydrodynamic timescales
and therefore account for the droplet's macroscopic behavior.

We extend the IBM to incorporate sound waves by identifying
the time-averaged acoustic radiation force on
each element of the droplet surface \cite{danilov2000mean, karlsen2015forces, silva2018acoustic, bruus2012acoustofluidics7, westervelt1957acoustic}:
\begin{equation}
    \label{eq:acousticradiationforce}
    \vec{F}^{ARF}_i(t)
    =
    -\AVE{p_2}  \mathbf{\hat{n}} - \rho_0\AVE{(\vec{u}_1\cdot\mathbf{\hat{n}})\, \vec{u}_1},
\end{equation}
where the continuous fields are evaluated at $\vec{x} = \vec{X}_i$ and
angle brackets represent an average
over one acoustic cycle.
The first term on the right-hand side of
Eq.~\eqref{eq:acousticradiationforce} describes acoustic
radiation pressure, and the second term arises from
advection of fluid by the droplet's oscillating boundary
\cite{danilov2000mean, karlsen2015forces}.

The net force on the boundary marker at $\vec{X}_i$,
\begin{equation}
    \vec{F}_i(t)
    =
    \vec{F}^{ST}_i(t)
    +
    \vec{F}^{ARF}_i(t),
\end{equation}
is the sum of the surface tension from Eq.~\eqref{eq:surfacetension}
and the time-averaged acoustic force
from Eq.~\eqref{eq:acousticradiationforce}.
The interfacial force drives streaming
flows,
$\AVE{\vec{u}(\vec{x}_j, t)} = \vec{u}_2(\vec{x}, t)$,
and shape deformations,
$\partial_t\vec{X}_i=\vec{U}_i$, through the cycle depicted in Fig.~\ref{fig:ibm}.
This extension to the incompressible IBM is one of the principal
contributions of the present work.
Its implementation requires expressions for the first- and second-order fields.

\subsection{First-Order Equations for the Acoustic Wave}
\label{sec:firstorderacoustics}

The first-order Navier-Stokes equations,
\begin{subequations}
\label{eq:pert1}
\begin{align}
    \partial_t\rho_1
    & =
    -\rho_0 \, \DIV{u}_1
    \label{eq:pert1a}\\
    \rho_0 \, \partial_t\vec{u}_1
    & =
    -\GRAD p_1
    + \eta \, (\beta + 1) \, \GRAD (\DIV{u}_1) \nonumber \\
    & \quad
    - \eta \, \GRAD \times \GRAD \times \vec{u}_1,
    \label{eq:pert1b}
\end{align}
\end{subequations}
can be simplified to a pair of wave equations,
\begin{subequations}
\label{eq:helm}
\begin{gather}
    \nabla^2 \phi
    =
    - \frac{k^2}{\gamma} \, \phi \label{eq:helma}\\
    \nabla \times \nabla \times \vec{\psi}
    =
    k_\nu^2 \vec{\psi}
    \label{eq:helmb},
\end{gather}
\end{subequations}
by introducing the Helmholtz decomposition,
\begin{subequations}
    \label{eq:helmholtz}
    \begin{gather}
        \vec{u}_1(\vec{r})
        =
        \GRAD\phi
        +
        \CURL{\psi}, \label{eq:helmholtza}\\
        p_1(\vec{r})
        =
        \frac{i\omega\rho_0}{\gamma} \, \phi(\vec{r}),
    \end{gather}
\end{subequations}
and noting that $p_1(\vec{r}) = c_0^2 \, \rho_1(\vec{r})$
in an isentropic fluid medium with speed of sound $c_0$.
Equation~\eqref{eq:helma} describes a scalar pressure wave
that propagates with the acoustic wavenumber, $k = \omega/c_0$.
Likewise, Eq.~\eqref{eq:helmb} describes vortical waves that carry away
the acoustic energy lost to viscous damping.
The acoustic damping coefficient,
\begin{equation}
      \gamma
      =
      1 + (\beta + 1) \left(\frac{k}{k_\nu}\right)^2 ,
\end{equation}
and the viscous wave number,
\begin{equation}
    k_\nu = \frac{1+i}{\delta}, \label{eq:viscouswavenumber}
\end{equation}
are both characterized by the thickness of the viscous boundary layer \cite{settnes2012forces},
\begin{equation}
    \delta(\omega) = \sqrt{\frac{2\eta}{\rho_0}\frac{1}{\omega}} .
\end{equation}

Following conventional acoustic radiation theory \cite{settnes2012forces, abdelaziz2021ultrasonic}, we neglect viscosity in the first-order fields, so that $\gamma \approx 1$ and $\vec{\psi} \approx \vec{0}$.
The solution to Eq.~\eqref{eq:helm}
can then be found by a multipole expansion of $\phi(\vec{r})$ both inside and outside the droplet \cite{silva2011expression},
\begin{subequations}
    \label{eq:multipole}
    \begin{align}
    \label{eq:multipolea}
    \phi_\text{I}(\vec{r})
        & =
        \phi_0\sum_{\ell=0}^\infty
        \sum_{m=-\ell}^\ell
        b_\ell^m \,
        j_\ell(k_\text{I} r) \,
        P_\ell^m(\cos\theta)
        \quad \text{and} \\
        \phi_\text{O}(\vec{r})
        & =
        \phi_0\sum_{\ell=0}^\infty
        \sum_{m=-\ell}^\ell
        a_\ell^m
        \left[
        j_\ell(kr)
        +
        s_\ell^m
        h_\ell(kr)
        \right]
        P_\ell^m(\cos\theta) ,
    \end{align}
\end{subequations}
respectively, where $j_\ell$
and $h_\ell$ are the spherical Bessel and Hankel functions, respectively,
and $P_\ell^m$ is the associated Legendre polynomial.
Distances in Eq.~\eqref{eq:multipolea}
are scaled by
the wavenumber inside the droplet, $k_\text{I} = \omega/c_\text{I}$.

The beam-shape coefficients, $a_\ell^m$, depend on the
structure of the incident sound wave, which typically is known \emph{a priori}.
The scattering coefficients, $s_\ell^m$ and $b_\ell^m$,
are obtained by satisfying boundary conditions at the droplet surface,
\begin{subequations}
    \label{eq:boundaryconditions}
    \begin{gather}
        p_\text{I}
        =
        p_\text{O} \quad \text{and} \\
        \vec{u}_\text{I}\cdot\mathbf{\hat{n}}
        =
        \vec{u}_\text{O}\cdot\mathbf{\hat{n}}.
    \end{gather}
\end{subequations}
For a spherical droplet of radius $R$, which is the simplest case,
the scattering coefficients
reduce to
\begin{subequations}
\label{eq:sphericalcoefficients}
    \begin{align}
        b_\ell^m
        & =
        \tilde{\rho}a_\ell^m
        \left[
        \frac{j_\ell(kR)}{j_\ell(k_IR)}
        +
        s_\ell^m \frac{h_\ell(kR)}{j_\ell(k_IR)}
        \right] \quad \text{and}
        \\
        s_\ell^m
        & =
        - \frac{
        \tilde{\rho}\tilde{c} \,
        j_\ell(kR) \,
        j_\ell^\prime(k_\text{I}R)
        -
        j_\ell(k_\text{I}R) \,
        j_\ell^\prime(kR)}
       {\tilde{\rho}\tilde{c} \,
        h_\ell(kR) \,
        j_\ell^\prime(k_\text{I}R)
        -
        j_\ell(k_\text{I}R) \,
        h_\ell^\prime(kR)} ,
    \end{align}
\end{subequations}
where
$\Tilde{\rho} = \rho_\text{O}/\rho_\text{I}$ and
$\Tilde{c} = k_\text{I}/k = c_\text{O}/c_\text{I}$, and primes denote derivatives with respect to the argument.
In the special case $\Tilde{c} = 1$, the scattering coefficients simplify further to
\begin{subequations}
\label{eq:simplifiedsphericalcoefficients}
    \begin{align}
        b_\ell ^m
        & =
        \tilde{\rho}a_\ell^m
        \left[
        1
        +
        s_\ell^m
        \frac{h_\ell(kR)}{j_\ell(kR)}
        \right] \quad \text{and}
        \\
        s_\ell ^m
        & =
        -\frac{
        \Tilde{\rho}
        -
        1
        }
       {\tilde{\rho}\frac{h_\ell(kR)}
        {j_\ell(kR)}
        -
        \frac{h_\ell^\prime(kR)}
        {j_\ell^\prime(kR)}} .
    \end{align}
\end{subequations}
Equations~\eqref{eq:sphericalcoefficients}
and \eqref{eq:simplifiedsphericalcoefficients}
serve as a point of departure for describing scattering by
aspherical droplets.

\subsection{Second-Order Equations for Time-Averaged Dynamics}

The first-order incident and scattered sound waves together give rise to steady forces and flows
at second order.
The steady, time-averaged second-order Navier-Stokes equations,
\begin{subequations}
\label{eq:pert2}
\begin{align}
      -\GRAD\cdot\AVE{\rho_1\vec{u}_1}
      & =
      \rho_0 \AVE{\GRAD \cdot\vec{u}_2} \label{eq:pert2a} \\
      \AVE{\rho_1\partial_t\vec{u}_1 +\rho_0(\vec{u}_1\cdot\GRAD)\vec{u}_1}
      & = -\GRAD \AVE{p_2}+\eta\nabla^2\AVE{\vec{u}_2} \nonumber \\
      & \quad + \beta\eta\GRAD\AVE{\DIV{u}_2} ,
      \label{eq:pert2b}
\end{align}
\end{subequations}
reduce to
\begin{align}
    -\vec{\nabla} \AVE{p_2} + \eta\GRAD\times\AVE{\CURL{u}_2}
    =
    \vec{f}_R
    \label{eq:streaming}
\end{align}
in the limit that that the fluid is incompressible, $\AVE{\DIV{u}_2} = 0$
\cite{nyborg1965acoustic}.
The Reynolds stress that drives the flow,
\begin{equation}
    \vec{f}_R(\vec{r}) \equiv \rho_0\GRAD\cdot \AVE{\vec{u}_1\vec{u}_1} ,
    \label{eq:reynolds}
\end{equation}
is obtained from the first-order solution. The acoustic radiation pressure, $\AVE{p_2}$, and vortical acoustic streaming, $\AVE{\CURL{u}_2}$, are given by the irrotational and solenoidal components of $\vec{f}_R$, respectively.
The latter term vanishes because we neglect viscosity
in the first-order fields, which means that $\vec{u}_1(\vec{r}) = \GRAD\phi$
and therefore that the Reynolds stress is irrotational.
Consequently, the viscous flows described by
the second term on the left-hand side of Eq.~\eqref{eq:streaming} vanish.
From this, we conclude that
the radiation pressure is \cite{bruus2012acoustofluidics2,bruus2012acoustofluidics7}
\begin{equation}
 \AVE{p_2} =
    \frac{1}{2} \rho_0\AVE{u_1^2}
    -
    \frac{1}{2} \kappa_0\AVE{p_1^2},
    \label{eq:acousticradiationpressure}
\end{equation}
where $\kappa_0 = (\rho_0 c_0^2)^{-1}$ is the isentropic compressibility of the fluid.
The second-order pressure wave
depends on the squares of the first-order
fields and therefore
does not vanish on an average over the
acoustic period.
It nevertheless can vary on
inertial time scales.

When the boundary layer is thin, $\delta(\omega) \ll R$, Eq.~\eqref{eq:acousticradiationpressure} is a good approximation for $\AVE{p_2}$ evaluated
at the time-averaged position of the droplet surface, and therefore
completes the description of the sound wave required to model
the acoustic force acting on the droplet through
Eq.~\eqref{eq:acousticradiationforce}.
While we have neglected viscous streaming due to the Reynolds stress,
$\vec{f}_R(\vec{r})$, in Eq.~\eqref{eq:reynolds},
our simulations still have acoustic streaming flows that are
driven by the the momentum flux advected by the boundary, which is described by the second term on the right-hand side of Eq.~\eqref{eq:acousticradiationforce}.

\subsection{Acoustic Radiation Force on an Ellipsoid}
\label{sec:ellipsoid}

Together, Eq.~\eqref{eq:acousticradiationforce}
and Eq.~\eqref{eq:acousticradiationpressure}
express the second-order acoustic radiation force on the droplet
in terms of first-order acoustic fields.
Analytic solutions for these first-order fields
are known for certain geometries with high symmetry.
More generally, the scattering coefficients in Eq.~\eqref{eq:multipole} may be obtained
via a variety of semi-analytic techniques.
These include the T-matrix method, which has been extended to acoustics by Waterman \cite{waterman2009t},
the discrete dipole approximation \cite{yurkin2007discrete}, and modal-matching methods \cite{mitri2015axisymmetric, mitri2017axial}. Here, we introduce an alternative approximation scheme
that leverages the exact spherical solution and
consequently is more efficient when deformations are small.

\subsubsection{Approximate solutions for slightly aspherical droplets}
\label{sec:aspherical}

\begin{figure}
    \centering
    \includegraphics[width=0.5\textwidth]{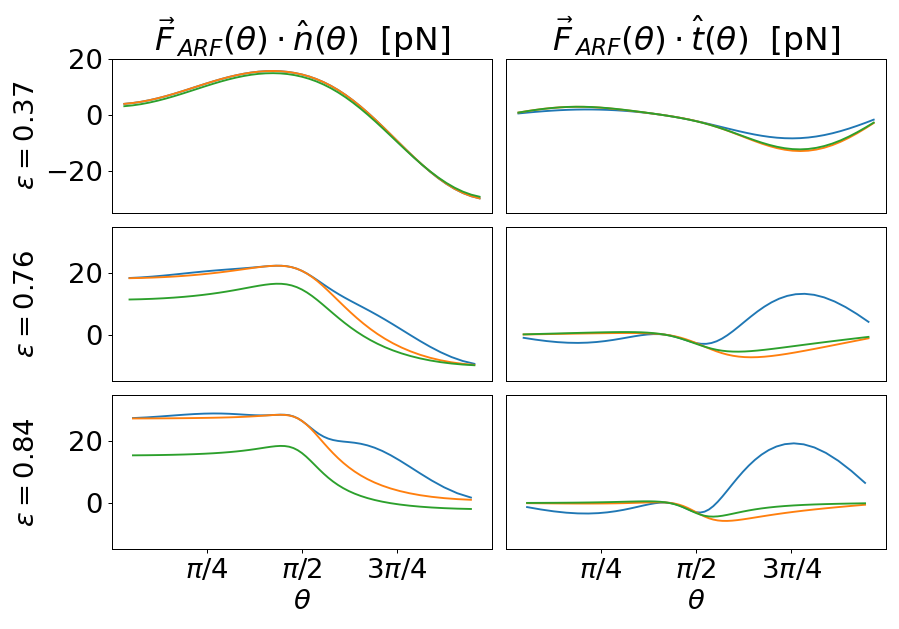}
    \caption{ARF profile from $\theta \in [0, \pi]$ for ellipsoids of various eccentricities at a displacement from the node of $z_0 =- \SI{0.6}{\um}$.
    The ARF at the surface is calculated using the spherical fields from
    Eq.~\eqref{eq:simplifiedsphericalcoefficients} (blue), the projection correction described in
    Eq.~\eqref{eq:projections} (orange), and the exact analytic solution in spheroidal coordinates (green).}
    \label{fig:scattering}
\end{figure}
To leading order, the fields scattered by a slightly aspherical droplet are approximated by the fields scattered by the minimally enclosing sphere.
The simple spherical approximation, however, does not satisfy the appropriate boundary conditions for the fluid velocity at the droplet's surface,
\begin{subequations}
    \label{eq:projections}
\begin{align}
    [\vec{u}_O(\vec{r}_s) - \vec{u}_I(\vec{r}_s)] \cdot \mathbf{\hat{n}}(\vec{r}_s)
    & = 0 \\
    [\Tilde{\rho} \, \vec{u}_O(\vec{r}_s) - \vec{u}_I(\vec{r}_s)] \cdot \mathbf{\hat{t}}(\vec{r}_s)
    & = 0 ,
\end{align}
\end{subequations}
where $\mathbf{\hat{t}}$ is the surface tangent unit vector,
and $\mathbf{\hat{t}} \times \mathbf{\hat{n}} = 0$.
Taking $\vec{U}(\vec{r})$ to be the flow field that would
be created by the enclosing
sphere,
the actual flow field at position $\vec{r}_s$ on the surface
of the distorted sphere can be approximated by
\begin{subequations}
\label{eq:approximations}
\begin{align}
    \vec{u}_I(\vec{r}_s)
    & =
    \vec{U}_I(\vec{r}_s), \\
    \vec{u}_O(\vec{r}_s)
    & =
    \vec{U}_O(\vec{r}_s) + \vec{\delta}(\vec{r}_s).
\end{align}
\end{subequations}
This flow field satisfies the boundary conditions in
Eq.~\eqref{eq:projections} if
\begin{subequations}
\label{eq:delta}
\begin{align}
    \vec{\delta}(\vec{r}_s) \cdot \mathbf{\hat{r}}
    & =
    n_\theta(\vec{r}_s)
    \left(1 - \frac{1}{\tilde\rho}\right)
    \,
    \vec{U}_I(\vec{r}_s) \cdot \mathbf{\hat{t}}(\vec{r}_s),
    \quad \text{and} \\
    \vec{\delta}(\vec{r}_s) \cdot \mathbf{\hat{\theta}}
    & =
    n_\theta(\vec{r}_s)
    \left(1 - \frac{1}{\tilde\rho}\right)
    \,
    \vec{U}_I(\vec{r}_s) \cdot \mathbf{\hat{n}}(\vec{r}_s),
\end{align}
where $n_\theta \equiv \mathbf{\hat{n}}(\vec{r}_s) \cdot \mathbf{\hat{\theta}}$.
\end{subequations}

The velocity field obtained with Eq.~\eqref{eq:delta} does not necessarily satisfy the Helmholtz equation,
nor is it necessarily curl-free.
Both of these conditions are \emph{approximately} satisfied, however,
so long as the droplet's distortions are small,
$n_\theta \ll 1$.
Moreover, $\vec{u}(\vec{r})$
satisfies the boundary conditions from Eq.~\eqref{eq:boundaryconditions},
and therefore should accurately represent the
induced flow near the droplet's surface.

\subsubsection{Comparison to analytic solution for an ellipsoid}

We test this approximation scheme on the special case of spheroidal droplets,
whose scattering coefficients can be expressed analytically in spheroidal coordinates
\cite{silva2011expression}.
The associated spheroidal wave functions are much more costly to evaluate than
spherical harmonics, which limits the utility of the analytic solution
for time-resolved simulations.
Analytic solutions for spheroids are useful, nonetheless, for validating our computationally
efficient approximation scheme.

Figure~\ref{fig:scattering} compares analytic
and approximated force profiles computed for
spheroids of various eccentricities.
Each spheroid is taken to be displaced by
$z_0 = \SI{-0.6}{\um}$ from the node of a standing wave;
the influence of the droplet's displacement is discussed further
in Appendix \ref{sec:applicability}.
The spherical solution, $\vec{u}_O(\vec{r}_s) = \vec{U}_O(\vec{r}_s)$, plotted in blue,
has qualitatively different behavior from the true solution, plotted in green,
even at small deformations.
By contrast, the corrected solutions, $\vec{u}_O(\vec{r}_S)=\vec{U}_O(\vec{r}_s) + \vec{\delta}(\vec{r}_s)$, plotted in orange, agrees with analytic radiation force for moderate deformations.
The normal component of the computed force tracks the functional form of the
exact solution up to a reasonably small multiplicative factor.
This discrepancy should not significantly affect the shape or dynamics of
the droplet, which is assumed to be incompressible.
Agreement with the analytic solution is much better for the tangential forces, which are responsible for driving acoustic streaming flows in our simulations.
The hybrid IBM therefore should predict the structure of streaming flows far more
accurately than the simple spherical approximation, and far more efficiently
than the analytic solution.
A further discussion of limitations of Eq.~\eqref{eq:delta} is provided
in Appendix~\ref{sec:applicability}.

\section{Results: Droplet in a Standing Plane Wave}
\label{sec:implementation}

Section~\ref{sec:ibm} introduces an IBM that models
the dynamics of a
droplet suspended in a fluid under
the general influence of external forces.
Section~\ref{sec:acoustic} explains
how sound propagating through the fluid generates such forces.
Purely analytic solutions to the sound-driven moving-boundary
problem are not yet available.
Any purely numerical description of the sound wave \cite{wang2020numerical,seo2011high} would have to
iterate through an enormous range of time scales to capture the moving droplet's steady dynamics.
Our hybrid implementation describes the sound wave semi-analytically
and handles the moving-boundary problem numerically.
The result is an accurate and exceptionally efficient
implementation.

We demonstrate the efficacy of the hybrid IBM
with an illustrative model system:
a single droplet levitated in a
planar acoustic standing wave.
To set up the IBM for this system, we specify the
standing wave's scalar potential,
\begin{equation}
    \phi_i(\vec{r})
    =
    \phi_0 \sin(kz),
\end{equation}
with amplitude $\phi_0$ and a nodal plane at $z=0$.
The associated beam-shape coefficients
\cite{silva2018acoustic},
\begin{equation}
    a_\ell^m = (2\ell+1)\cos\left(kz+\frac{\ell\pi}{2}\right) .
\end{equation}
yield a compact expression for the the net ARF on the droplet:
\begin{equation}
    \sum_{i=0}^{N_b} \vec{F}_i^{ARF}
    =
    \frac{1}{2} \rho_0 \phi_0^2 Q \sin(2 kz) \, \hat{z},
\end{equation}
where the radiation force efficiency, $Q$, is a dimensionless
scattering coefficient that depends on droplet geometry and boundary conditions \cite{silva2018acoustic,abdelaziz2021ultrasonic}. For a rigid sphere in a standing wave, $Q=k^3 V_p/5$ \cite{silva2018acoustic}, and $V_p$ is the volume of the particle.
The scale of the acoustic force,
$\frac{1}{2} \rho_0 \phi_0^2$, is set by the driving voltage in
experimental realizations \cite{abdelaziz2021ultrasonic}.

The simulations described in this section all
start with a spherical droplet of radius $R = \SI{2}{\um}$
and density $\rho_I = \SI{1.18}{\g\per\cubic\cm}$, dispersed in water.
The droplet is subject to
a buoyant force,
$F_g = (\rho_I - \rho_0) \, V_p g \approx \SI{60}{\pico\newton}$, that
is balanced by acoustic forces.
To model practical acoustic levitators for water-borne samples \cite{abdelaziz2021ultrasonic},
the driving frequency is set to \SI{2}{\mega\hertz}.
The dimensionless size parameter, $kR = \num{0.034}$, is well within
the Rayleigh limit, $kR \ll 1$.
The droplet initially is released at
the nodal plane, and then relaxes to its equilibrium
position, $z_0$, and to its equilibrium shape.

\subsection{Separation of Timescales}
\label{sec:timescales}

Modelling a levitated droplet
illustrates the advantages of our hybrid method relative to conventional numerical techniques.
The acoustic IBM presented in Ref.~\cite{wang2020immersed}, for example,
computes the influence of acoustic forces by directly solving
the compressible Navier-Stokes equations using a 5th-order
Weighted Essentially Non Oscillatory (WENO) scheme.
When applied to a system that is comparable to ours \cite{wang2020numerical},
the WENO scheme numerically resolves acoustic wave propagation
with timestep $dt \approx R/c \approx \SI{1.33}{\ns}$.
By treating acoustic-scale processes semi-analytically, the hybrid IBM
can resolve the droplet's dynamics with time steps of \SI{15}{\us},
which represents an inherent ten-thousand-fold improvement in efficiency.
\begin{figure*}
    \centering
\includegraphics[width=\textwidth]{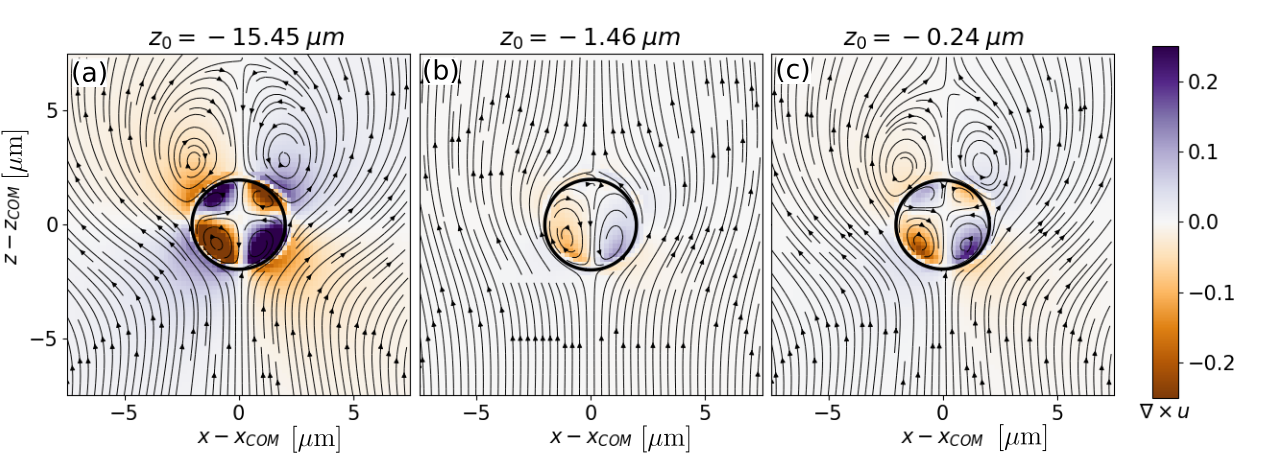}
    \caption{Streaming profiles for a droplet levitated
    in water with $\sigma = \SI{3.3}{\nano\newton\per\um}$
    and $R = \SI{2}{\um}$ at various trap strengths.
    (a) Gravity displaces the droplet downward from the nodal
    plane, favoring quadrupolar streaming in the weakest
    levitator. (b) Increasing trap strength induces a
    crossover to dipolar streaming. (c) Increasing
    trap strength further induces a second crossover
    to quadrupolar flow. The flow's vorticity is indicated
    by color.}
    \label{fig:var_pos}
\end{figure*}

\begin{figure*}
    \centering
    \includegraphics[width=\textwidth]{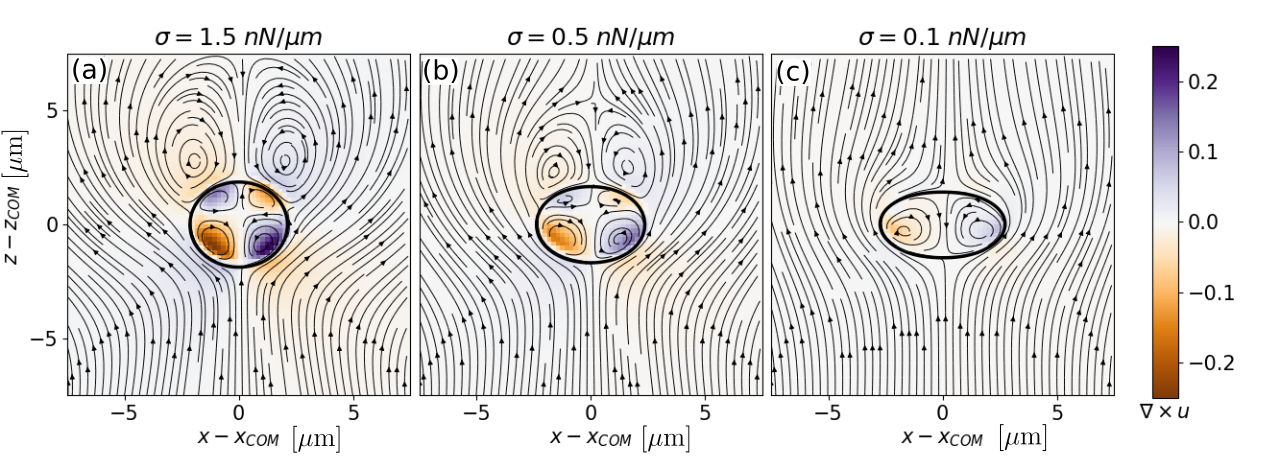}
    \caption{Streamlines of the streaming flow for
    a droplet of radius $R = \SI{2}{\um}$ in a strong
    levitator, $z_0 \approx \SI{-0.15}{\um}$.
    (a) Reducing the surface tension, $\sigma$,
    by half relative
    to Fig.~\ref{fig:var_pos} allows the droplet to
    deform while retaining quadrupolar
    flows.
    (b) Further reducing surface tension increases
    the droplet's distortion while significantly
    decreasing the vorticity in the streaming flow.
    (c) A highly deformable droplet induces dipolar
    streaming flows.}
    \label{fig:var_stiff}
\end{figure*}

\subsection{Equilibrium Shape of a Levitated Droplet}
\label{sec:equilibriumshape}

\begin{center}
\begin{figure}
    \includegraphics[width=1.1\columnwidth]{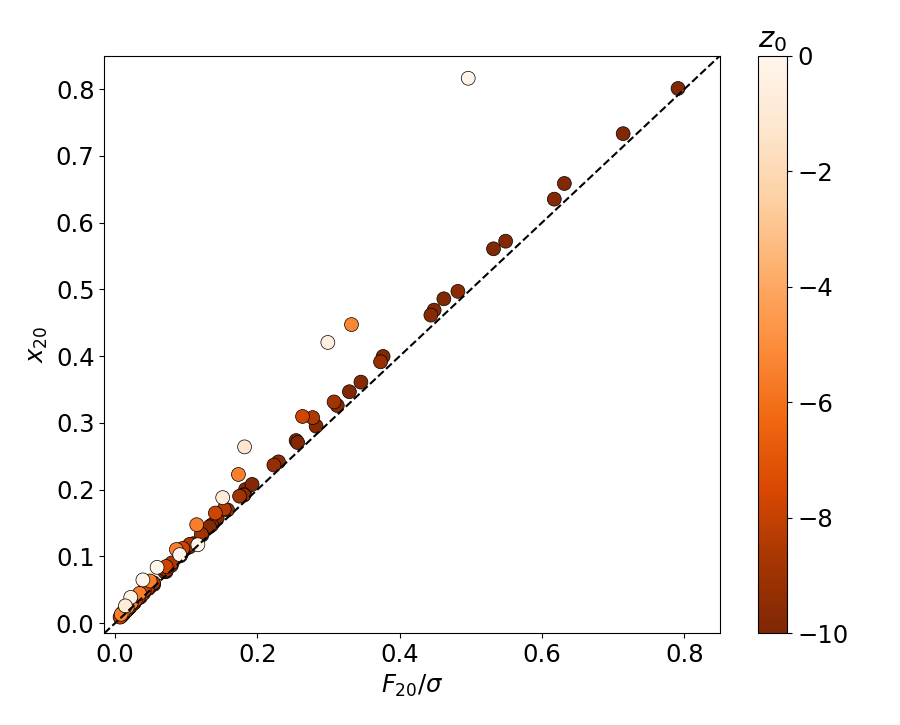}
    \caption{Dependence of a droplet's degree of deformation on the relative strength of the acoustic driving. Outliers represent droplets that were not stably trapped.}
    \label{fig:response}
\end{figure}
\end{center}

The equilibrium shape of a droplet in an acoustic field
is typically found analytically \cite{trinh1986equilibrium,jackson1988equilibrium}
by relating the surface curvature to acoustic stresses
with the Young-Laplace formula.
Working in two dimensions, the equilibrium
shape of a slightly deformed droplet
of radius $R$
may be expanded in associated Legendre polynomials
\cite{marston1980shape},
\begin{equation}
    r(\theta)
    =
    R
    +
    \sum_{\ell = 0}^\infty
    \sum_{m=-\ell}^\ell
    x_{\ell m} \,
    P_\ell^m(\cos\theta).
\end{equation}
When the droplet has the same viscosity as the medium,
the normal force, $F_n(\theta) = \vec{F}^{ARF}\cdot\hat{n}$,
determines the static shape deformation \cite{marston1980shape,marston1981quadrupole},
\begin{equation}
    \label{eq:response}
    x_{\ell m}
    =
    \frac{R^2}{(\ell + 2)(\ell - 1) \, \sigma}
    \int_{-1}^1 F_n(\theta) \, P_\ell^m(\cos \theta) \, d\cos\theta.
\end{equation}

In a standing wave, the primary deformation mode
is described by $x_{20}$, which flattens the droplet
into a roughly elliptical shape.
Figure~\ref{fig:response} compares the value for $x_{20}$ observed in
IBM simulations of a droplet of $R = \SI{2}{\um}$ with predictions of Eq.~\eqref{eq:response}
over a range of values of surface tension,
$\sigma$, and trap strength, $\phi_0^2$.
The computed deformation agrees well for droplets close to the trapping plane, and deviates at larger displacements from the node. This is because our approximation for scattering, Sec.~\ref{sec:aspherical}, begins to break down at larger displacements from the node.

\subsection{Acoustic Streaming }
\label{sec:sphericalstreaming}

Acoustic forces not only translate and deform the droplet, but also generate interior and exterior streaming flows. In the case of a spherical droplet, these flow fields can be described analytically. The streaming profiles for an acoustically levitated droplet are characterised by a transition from dipolar flow to quadrupolar
flow depending on the relative magnitude of acoustically-driven pulsation and translation modes respectively \cite{baasch2020acoustic}. The streaming behavior depends on properties of the droplet and surrounding fluid, and can also be manipulated directly by adjusting the sound intensity, $\phi_0^2$, of the trap to control the droplet's position relative to the node.
\subsubsection{Spherical Droplet}
The streaming profiles around a spherical droplet have been computed analytically by
Baasch, Doinikov, and Dual in Ref.~\cite{baasch2020acoustic}.
Their solution should be comparable
to results of the hybrid IBM under conditions where
shape deformations are small,
\emph{i.e.} $\sigma R \gg \rho_0\phi_0^2$.
The results in Fig.~\ref{fig:var_pos} are obtained by fixing the surface tension at a fairly large value, $\sigma = \SI{3.3}{\nano\newton\per\um}$,
and varying the trap strength, $\phi_0^2$.
Because the droplet is more dense than its medium, it sinks
below the nodal plane by a distance that is inversely proportional to the
traps' strength and therefore
samples different regions of the acoustic force landscape.
This has consequences for the nature of the induced
streaming flow.

Figure~\ref{fig:var_pos}(a) shows the quadupolar streaming flow
that is induced when the droplet is far below the nodal plane
in a comparatively weak acoustic trap.
Increasing $\phi_0$ by a factor of \num{3} in Fig.~\ref{fig:var_pos}(b)
lifts the droplet toward the nodal plane and qualitatively transforms
the streaming flow into a principally dipolar motif.
This transition from quadrupolar to dipolar flow has been
with increasing trap stiffness has been predicted analytically
and is reported in
Fig.~7 of Ref.~[\onlinecite{baasch2020acoustic}].
Further increasing the trap stiffness in Fig.~\ref{fig:var_pos}(c)
lifts the particle still closer to the nodal plane and induces
second transition back to quadrupolar streaming
that appears not to have
been reported previously.

\subsubsection{Deformable Droplets}

Both dynamical transitions reported in Fig.~\ref{fig:var_pos}
occur for droplets that are stiff enough to remain
substantially spherical while trapped.
This is consistent with the assumption of sphericity
that underlies the analysis in Ref.~\cite{baasch2020acoustic}.
Figure~\ref{fig:var_stiff} reveals
yet another dynamical transition that occurs when a strongly-trapped
droplet is soft enough to deform,
a scenario that has not been considered in previous studies.
The droplet in Fig.~\ref{fig:var_stiff}(a) is comparable to the
droplet in Fig.~\ref{fig:var_pos}(c), except that its surface
tension is smaller by a factor of two.
The droplet consequently deforms into an ellipsoid under the uniaxial
stress of the acoustic levitator and settles at a slightly
different height relative to the nodal plane.
The induced streaming flow nevertheless retains its quadrupolar
nature.

Reducing the surface tension, in Fig.~\ref{fig:var_stiff}(b),
increases the distortion and moves both the interior and exterior circulation closer to the droplet surface.
Reducing the surface tension still further, in Fig.~\ref{fig:var_stiff}(c),
suppresses the vortex structure entirely. The streaming flow around
the ellipsoid droplet is primarily dipolar.
This deformation-induced dynamical transition also appears not
to have been observed in prior studies.
These observations illustrate the value of the hybrid IBM for
probing the dynamical properties of insonated droplets.

\section{Discussion}
\label{sec:dicussion}

We have introduced a hybrid immersed boundary method that efficiently models the stresses on acoustically levitated droplets and the surrounding fluid.
We have validated our method by simulating a single droplet in a plane standing wave, and comparing the results with the best available analytical and numerical solutions.
Even in this minimal system, the hybrid IBM reveals
transitions between dipolar and quadrupolar streaming
under strong-driving conditions
that have not been reported in previous studies.
These observations offer new insights into the response
of deformable droplets to the forces and torques created
by acoustic landscapes.

Compared to traditional compressible flow simulations, our hybrid method cuts computational cost by utilizing semi-analytic techniques to accommodate acoustic scattering in hydrodynamic simulations.
The benefits are greatest when stress-induced deformations are small
enough that analytic scattering calculations can be carried out
rapidly.
When the droplet shape is highly irregular and unpredictable,
most known semi-analytic techniques converge slowly and may be prohibitively expensive.
However, for droplets with small to moderate deformations, the hybrid IBM may be up to three orders of magnitude faster than compressible flow simulations. For the implementation presented here, the hybrid method efficiently and effectively models complex shape deformations and streaming profiles around a single droplet, and can easily be scaled up to study acoustohydrodynamic interactions among multiple particles.

The framework presented here is, by design, extremely flexible. If necessary, the accuracy of the hybrid IBM can be improved by introducing more advanced scattering solutions for Eq.~\eqref{eq:multipole}, implementing viscous waves from Eq.~\eqref{eq:helmb}, or even adding additional pairwise acoustic forces, like the Bjerknes force \cite{abdelaziz2021ultrasonic}. Or, as in this work, inexpensive approximations can be used throughout to efficiently model qualitative behaviors.

\section*{Acknowledgments}

This work was supported by the National Science Foundation under
Award Number DMR-2104837.

We are grateful to professor Charles Peskin for extensive discussions on the formulation of the Immersed Boundary Method.

\section*{Data Availability}

The data that support the findings of
this study are available from the
corresponding author upon reasonable
request.

\begin{figure*}
    \centering
    \includegraphics[width=\textwidth]{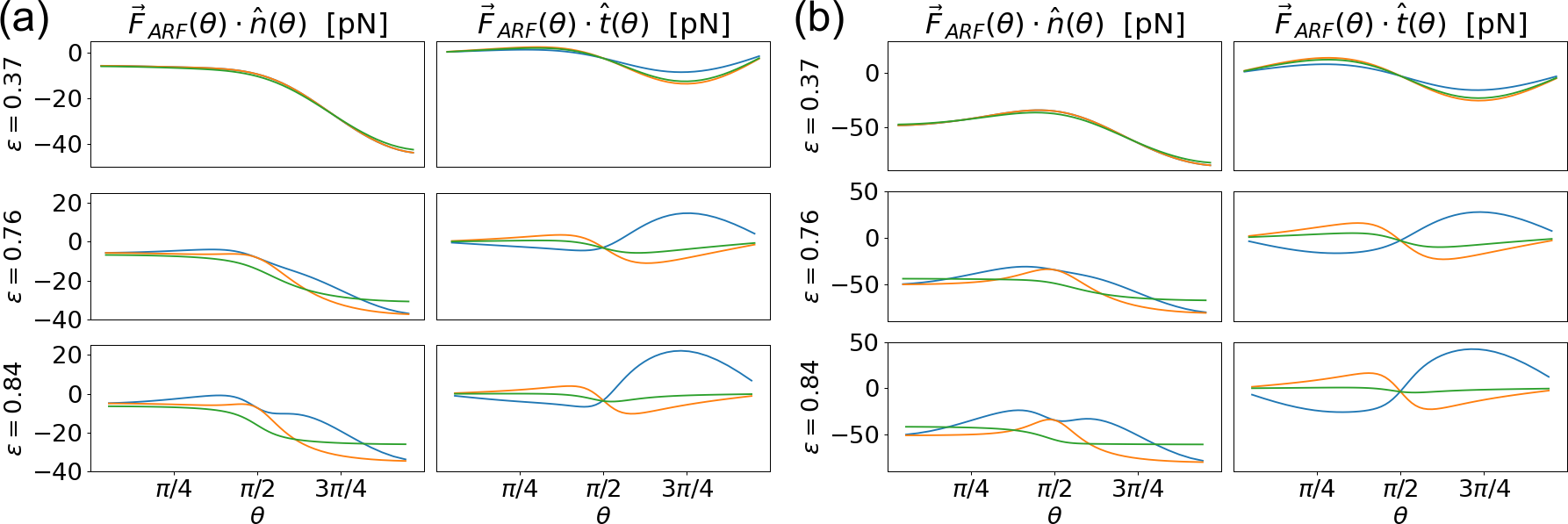}
    \caption{Normal and tangential components of the acoustic radiation force as a function of position, $\theta$,
    on the droplet surface for ellipsoids of various eccentricities.
    (a) moderate displacement: $z_0 = \SI{-3.5}{\um}$.
    (b) larger displacement: $z_0 = \SI{-10.6}{\um}$.
    The ARF at the surface is calculated using the spherical fields from
    Eq.~\eqref{eq:simplifiedsphericalcoefficients} (blue), the projection correction described in
    Eq.~\eqref{eq:projections} (orange), and the exact analytic solution in spheroidal coordinates (green).
    }
    \label{fig:forceerror}
\end{figure*}

\appendix
\section{Flows around a sedimenting
spherical droplet}
\label{sec:sedimenting}

Analytic solutions are available for the
viscous flow fields inside and around a
spherical fluid droplet as it moves with
with velocity $\vec{v} = v \hat{z}$ through
an immiscible fluid \cite{happel1983low}.
Assuming that the two fluids have
the same viscosity,
the stream functions inside and
outside the droplet are
\begin{align}
    \psi^{(i)}(\vec{r})
    & =
    \frac{1}{8} v \, r^2
    \left(\frac{r^2}{a_p^2} - 1
    \right) \sin^2\theta,
    \quad \text{and} \\
    \psi^{(o)}(\vec{r})
    & =
    \frac{1}{8} v \frac{a_p^3}{r}
    \left( 1 - 5 \frac{r^2}{a_p^2}
    \right) \sin^2\theta,
\end{align}
respectively. The associated flow field
is
\begin{equation}
    \vec{u}(\vec{r}) = \GRAD\times(\psi \hat{z}).
\end{equation}

\section{Applicability of the Scattering Approximation from Sec.~\ref{sec:ellipsoid}}
\label{sec:applicability}

The approximation in Sec.~\ref{sec:ellipsoid} is derived by assuming that the acoustic fields at the droplet surface, $r=S(\theta)$, are similar to that at a minimally enclosing sphere,
$r = R \equiv \text{max}[S(\theta)]$.
The interior potential from Eq.~\eqref{eq:multipolea} can be rewritten by Taylor expanding the radial function to 1st order in $S-R$,
\begin{align*}
    j_\ell(kS)
    & =
    j_\ell(k_\text{I} R)
    +
    (S-R) k j_\ell^\prime(k_\text{I} R)
    +
    \mathcal{O}\{(S-R)^2\}
\end{align*}
In the long-wavelength limit, $kR \ll 1$, we can use the small-argument asymptotic form $j_\ell(z) \propto x^\ell$, and so we have
\begin{align*}
    j_\ell(kS)
    &=
    \left[
    1 + \ell\left(\frac{S}{R}-1\right)
    \right] j_\ell(k_\text{I} R)
    +
    \mathcal{O}\{(S-R)^2\}
\end{align*}
so that $\phi_I(R) \approx \phi_I(S)$
as long as \begin{equation*}
    \left(\frac{S}{R}-1\right)\ll \frac{1}{\ell}
\end{equation*}
for all non-negligible modes.
Likewise, repeating this procedure for the scattered field reveals that $\phi_s(R) \approx \phi_s(S)$ when$ \left(\frac{S}{R}-1\right)\ll 1/(\ell+1)$.

Because the error in the radial function is also proportional to $\ell$, higher modes are approximated more accurately than lower modes. Consequently, the applicability of this method also depends on the local structure of the acoustic field. In this work, we have shown in Fig.~\ref{fig:scattering} that our approximation performs well even at moderately large deformations, when the droplet is close to the node. However, as illustrated by Fig.~\ref{fig:forceerror}, the range of validity is restricted to smaller deformations when the droplet is farther away from the node.

%

\end{document}